\documentclass[12pt]{article}
\usepackage{graphicx}
\usepackage{amsfonts,amssymb,amsmath}
\usepackage{multirow}
\usepackage{color}
\usepackage{slashed}
\newcommand{\eea}{\end{eqnarray}}

\newcommand{\ba}{\begin{array}}
\newcommand{\ea}{\end{array}}
\newcommand{\beq}{\begin{equation}}
\newcommand{\eeq}{\end{equation}}
\def\bt{\begin{table}}
\def\et{\end{table}}
\def\bc{\begin{center}}
\def\ec{\end{center}}
\def\bi{\begin{itemize}}
\def\ei{\end{itemize}}
\def\bea{\begin{eqnarray}}
\def\eea{\end{eqnarray}}
\def\beas{\begin{eqnarray*}}
\def\eeas{\end{eqnarray*}}
\def\Zp{Z^{\prime}}
\def\zp{z^{\prime}}

\begin{document}
\setcounter{page}{0}
\thispagestyle{empty}
\vspace*{-1.5cm}
\begin{flushright}
OSU-HEP-10-06\\
FERMILAB-PUB-10-195T\\
HD-THEP-10-12
\end{flushright}
\vspace{0.5cm}

\begin{center}
{\Large {\sc Hidden Extra $U(1)$ at the Electroweak/TeV Scale} }

\vspace*{1.5cm}
    B.~N.~Grossmann\footnote{Email address: benjamin.grossmann@okstate.edu}$^{\dag}$,
    B.~McElrath\footnote{Email address: bob.mcelrath@cern.ch}$^{\ddag}$,
    S.~Nandi\footnote{Email address: s.nandi@okstate.edu}$^{\dag*}$,
    and Santosh~Kumar~Rai\footnote{Email address: santosh.rai@okstate.edu}$^\dag$

\vspace*{0.5cm}
$^{\dag}$\textit{Department of Physics and Oklahoma Center for High Energy Physics,\\
Oklahoma State University, Stillwater, OK 74078, USA}\\
%$^{\ddag}${\it Theory Division, CERN, CH-1211, Geneva 23, Switzerland \\}
$^{\ddag}$\textit{Universit\"at Heidelberg, 16 Philosophenweg, 69120 Heidelberg, Germany}\\
$^*$\textit{Fermi National Accelerator Laboratory, P.O. Box 500, Batavia, IL 60510, USA\footnote{S. Nandi was a Summer Visitor at Fermilab}}
\end{center}

\vspace*{0.3in}
\begin{abstract}

We propose a simple extension of the Standard Model (SM) by adding
an extra $U(1)$ symmetry which is hidden from the SM sector. Such a
hidden $U(1)$ has not been considered before, and its existence at
the TeV scale can be explored at the LHC.
%\FIXME{Why?}
This hidden $U(1)$ does not couple directly to the SM particles, and
couples only to new $SU(2)_L$ singlet exotic quarks and singlet
Higgs bosons, and is broken at the TeV scale. The dominant signals
at the high energy hadron colliders are multi lepton and multi
$b$-jet final states with or without missing energy. We calculate
the signal rates as well as the corresponding Standard Model
background for these final states. A very distinctive signal is 6
high $p_T$ $b$-jets in the final state with no missing energy. For a
wide range of the exotic quarks masses
%and the $\Zp$ \CRed{(should we put $\Zp$ here as
%we have not considered its production cross section or signal)},
the signals are observable above
the background at the LHC.
\end{abstract}

%\FIXME{Need discussion of gauge anomalies}
%\FIXME{More discussion of motivation needed}

%\vspace*{0.0in}
PACS:$12.60.Cn, 12.60.Fr, 12.90.+b$
%--------------------------------------------------------------------------
\vfill
%\newpage

\section{Introduction} \label{sec:intro}

The Standard Model (SM) is now well established as the effective
theory of the quarks, leptons and the gauge bosons below the TeV
scale. However, it is almost universally believed that the SM is not
the final theory. Many extensions of the SM have been proposed
%\CRed{(proposed/motivated)}
to solve some of the shortcomings of the
SM. Grand unified theories have been proposed to unify the three
gauge couplings into one.
%\FIXME{ref}
Supersymmetric extensions have been
proposed to solve the gauge hierarchy problem.
%\FIXME{ref}
A singlet Higgs
boson, with a $Z_2$ symmetry has been added to the SM which can
serve as a plausible candidate for dark matter
\cite{Davoudiasl:2004be}. One or more extra space-like compact
dimensions has been added to the usual four dimensions to
incorporate TeV scale as the fundamental scale of gravity
\cite{Antoniadis:1998ig}, or to unify the gauge and Higgs bosons,
and as well as the fermions, and understand the Yukawa interactions
as part of the gauge interactions \cite{Gogoladze:2003ci}. Most of
these extensions involve new gauge interactions, commonly an extra
$U(1)$, as well as new particles beyond the SM.

Many kinds of extra $U(1)$ gauge symmetries have been considered.
These include the left-right symmetric model \cite{Pati:1974yy},
$SO(10)$ or $E_6$ GUTs, superstring $E_6$ models \cite{rizzo},
topflavor models \cite{topflavor}, and string-inspired supersymmetric
models.\cite{Han:2004yd}
In most of these models, the SM fermions and the Electroweak (EW)
Higgs boson carry non-trivial charges under the $U(1)$. Other
variations of the extra $U(1)$ symmetry, such as a hadro-phobic
$U(1)$, lepto-phobic $U(1)$, and an extra $U(1)$ which couples only
to the third family of fermions have been considered \cite{rizzo}.

Hidden sectors of matter are ubiquitous among models due to the need to
break supersymmetry, as well as the common addition of particle dark
matter which cannot be charged under the Standard Model gauge groups
unless it is very heavy.\cite{Bertone:2004pz}  An extra $U(1)$ can be
a natural way to link the ``dark'' sector with the Standard Model.

In this work, we consider an extra $U(1)$ symmetry \cite{li} in
which the SM particles (the SM fermions, gauge bosons and the EW
doublet Higgs bosons) are neutral. We call this a hidden extra
$U(1)$ \cite{li}. Only new exotic quarks, and the EW singlet Higgs
bosons couple to this extra $U(1)$ gauge boson. These exotic quarks
and the EW singlet Higgs bosons act as messenger particles between
the hidden $U(1)$ sector and the SM sector. This extra $U(1)$
symmetry is broken at the EW scale by the vacuum expectation value
(VEV) of the EW singlet Higgs boson. Thus this extra gauge boson,
the exotic quarks, and the singlet Higgs boson all acquire masses at
the EW scale, and can be searched for at high energy colliders, such
as the Tevatron and the LHC. The dominant signals of our scenario at
the hadron colliders are multi-$b$ and multi-lepton final states, with
little or no missing energy.

\section{The Hidden $U(1)$ Model} \label{sec:model}

Our gauge symmetry is the usual Standard Model  $SU(3)_C \times
SU(2)_L \times U(1)_Y$ supplemented by an extra $U(1)$ symmetry,
which we call $U(1)^\prime$. We introduce two exotic quarks $D_L$
and $D_R$ which are color triplets and singlets under the $SU(2)_L$
gauge symmetry, but charged under the $U(1)_Y$. We denote the gauge
boson for the $U(1)^\prime$ by $\Zp$. We also introduce a complex
Higgs field $S$ which is a color and EW singlet, and has a charge
$q^\prime$ under the $U(1)^\prime$. This singlet Higgs field has a
VEV $v_S$ at the TeV or EW scale, and breaks the $U(1)^\prime$
symmetry.

The Lagrangian for the gauge part of the interaction for the exotic
$D$ quark is given by the usual gauge interaction under the
$SU(3)_C$ symmetry with the gauge coupling $g_3$. The EW and
$U(1)^\prime$ interactions of the matter fields with the gauge bosons
can be obtained from the Lagrangian:
\begin{align}
\label{kinetic} \mathcal{L}&=
\overline{q}_L^i i\slashed{\mathcal{D}}_2 q_L^i +
\overline{\ell_L}^i i\slashed{\mathcal{D}}_2 \ell_L^i +
\overline{u}_{R}^i i\slashed{\mathcal{D}}_1 u_{R}^i +
\overline{d}_{R}^i i\slashed{\mathcal{D}}_1 d_{R}^i +
\overline{e}_R^i i\slashed{\mathcal{D}}_1 e_R^i +
\overline{D}i\slashed{\mathcal{D'}}_1 D
\end{align}
where $q_L$, $\ell_L$ are the $SU(2)_L$ quark and lepton doublets
while $u_{R}$, $d_{R}$, $e_R$, and $D$ are the $SU(2)_L$ up-,
down-quark, lepton and exotic quark singlets, respectively. The
different covariant derivatives are defined as
%\FIXME{$g^\prime$ notation not consistent.  How about $g_Y$, $g^\prime$ and $Y$, $Y^\prime$ instead?}
\begin{align}
\label{covariant}
\begin{split}
{\mathcal{D}}_{2\mu} &= \partial_\mu - i \frac{g_2}{2} \tau\cdot W_\mu - i \frac{g^\prime}{2} Y B_\mu,\\
{\mathcal{D}}_{1\mu} &= \partial_\mu - i \frac{g^\prime}{2} Y B_\mu,\\
{\mathcal{D^\prime}}_{1\mu} &= \partial_\mu - i \frac{g^\prime}{2} Y B_\mu - i g_{\zp} Y_{\zp} \Zp{_\mu},
\end{split}
\end{align}
where $\tau$'s are the Pauli matrices; $Y$, $Y_{\zp}$ are the charges of the matter fields under
the $U(1)_Y$ and the new gauge group $U(1)^\prime$ respectively; while $\Zp$ represents the
new gauge boson.

The Higgs potential, with the usual doublet Higgs $H$ and the EW singlet Higgs $S$, is
\begin{align}
\label{pot}
V(H,S) &= -\mu_H^2 (H^\dagger H) - \mu_S^2 (S^\dagger S) + \lambda_H (H^\dagger H)^2 +  \lambda_{HS} (H^\dagger H)
(S^\dagger S) + \lambda_S (S^\dagger S)^2.
\end{align}
We can also write a mass term for the vector-like quark,
\begin{align}\label{Dmass}
\mathcal{L}_{\textrm{mass}} &= M_D \overline{D}_L D_R.
\end{align}
Note that the new exotic vector-like quark $D$ is like a new flavor,
and it has color, hypercharge, and an extra $U(1)^\prime$
interaction, but no $SU(2)_L$ interaction. Since this new $D$ quark
is vector-like with respect to both $U(1)$ as well as $U(1)'$, the
model is anomaly free.  Without any other interaction, the $D$ quark
will be stable. As none of the SM particles are charged under the
new $U(1)^\prime$ symmetry, the new symmetry will remain hidden from
the SM, provided the gauge-kinetic-mixing terms are strongly
suppressed. However, its gauge quantum numbers allow flavor changing
Yukawa interactions with the bottom, strange, and down quarks via
the singlet Higgs boson $S$.
\begin{align} \label{Yextra}
\mathcal{L}_{\textrm{Extra Yukawa}} &= Y_{Db} \overline{D}_L b_R S
+ Y_{Ds} \overline{D}_L s_R S + Y_{Dd} \overline{D}_L d_R S + h.c.
\end{align}
Note that in order the above Lagrangian to be hypercharge singlet,
the hypercharge of both $D_L$ and $D_R$ must be equal to that of $b_R$.
This also requires that the $U(1)^\prime$ charge ($Y_{\zp}$) for the
exotic quark $D$ must satisfy $Y_{\zp}=q^\prime$. Such a term in the
Lagrangian leads to mixing between the down-type quarks with the new exotic
vector-like quark $D$, giving rise to EW decay modes for the heavy quark.

We assume that the parameters in the Higgs potential are such that
$H$ has  VEV at the electroweak (EW) and $S$ has a VEV around the
TeV scale. Then, in the unitary gauge, we can write the $H$ and $S$
fields as
%%%%
\begin{align}
\label{vevs}
H(x)&=\frac{1}{\sqrt{2}} \left(\begin{matrix}0 \\ v_H + H_0(x)\end{matrix}\right),&
S(x)&=\frac{1}{\sqrt{2}}(v_S + S_0(x)),
\end{align}
%%%%
where $v_H$ is the VEV of the doublet Higgs, and $v_S$ is the singlet
VEV. From the minimization of the Higgs potential, we obtain,
\begin{align}
\label{vevsq}
v_H^2 &= \frac{\mu_H^2 - \frac{\lambda_{HS}}{{2\lambda_S}} \mu_S^2}
{\lambda_H - \frac{\lambda_{HS}^2}{4\lambda_S}} , &
v_S^2 &=\frac{\mu_S^2 - \frac{\lambda_{HS}}{{2\lambda_H}} \mu_H^2}
{\lambda_S - \frac{\lambda_{HS}^2}{4\lambda_H}} .
\end{align}
The scalar mass-squared matrix in the ($H_0,S_0$) basis is given by
\begin{align}
\label{hmass}
{\cal{M}}^2 &=\left(\begin{matrix}
2\lambda_H \, v_H^2     &   \lambda_{HS} \, v_H \, v_S  \\
\lambda_{HS} \, v_H \, v_S &    2\lambda_S \, v_S^2
\end{matrix}\right).
\end{align}
The masses of the two mass eigenstate Higgs scalars $\phi_H$ and $\phi_S$
as well as their mixing angle $\beta$, in terms of the fundamental
parameters of the Lagrangian, can be obtained from the above mass
matrix. In particular, the mixing angle $\beta$ is given by
\begin{align}
\label{hmix}
\tan 2\beta &= \frac{\lambda_{HS} v_H v_S}{\lambda_S v_S^2 - \lambda_H v_H^2} \,\,\,.
\end{align}
In addition to the usual gauge interaction for the $H_0$ and $S_0$, the
interaction among the Higgs from the potential $V(H_0,S_0)$ after symmetry breaking is given by
\begin{align}
\begin{split}
V(H_0,S_0) &= \lambda_H v_H {H_0}^3
        + \frac{\lambda_{HS} v_S}{2}\left({H_0}^2 S_0 + H_0 {S_0}^2\right)
        + \lambda_S v_S {S_0}^3\\
        &\qquad + \frac{\lambda_H}{4} {H_0}^4 + \frac{\lambda_{HS}}{4} {H_0}^2 {S_0}^2 + \frac{\lambda_S}{4} {S_0}^4.
\end{split}
\end{align}
The interaction among the Higgs mass eigenstates $\phi_H$ and $\phi_S$
can be obtained by using
\begin{align}\label{hstates}
H_0 &= \phi_H \cos\beta + \phi_S \sin\beta , &
S_0 &= - \phi_H \sin\beta + \phi_S \cos\beta.
\end{align}

To explore the phenomenological implications of the model, we need
to consider the various mixings which lead to the effective
interaction of these exotic particles to SM particles and are
responsible for their decays. We have already considered the mixing
in the scalar sector of the model which has interesting consequences
for Higgs searches at colliders such as LHC
\cite{Barger:2009me}.
%\FIXME{Nothing mentioned about Higgs searches -- this deserves its own section, or just some references e.g. Gunion et.al. higgs-to-higgs decays}
We also find that by allowing the Yukawa interactions given in
Eq.~\ref{Yextra}, there will be mixing between the down-type quarks
with the new exotic quark $D$ once the scalar $S$ gets a VEV. The
mixing between the down-type quarks with the exotic $D$ quark gives
rise to EW decay modes for the heavy quark. The heavy $Z^\prime$
also has additional interactions which lead to interesting decay
modes.

For simplicity, we assume that only $Y_{Db}$ is non-zero in
Eq.~\ref{Yextra} while the other Yukawa coefficients are negligibly
small. This would imply that the exotic quark mixes only with the
bottom quark, thus indirectly affecting the top-bottom vertex
($V_{tb}$ in the CKM matrix) as well as inducing a coupling between
the exotic $D$ quark with the top quark. The mixing can be
parametrized in terms of two mixing angles $\theta_L$ and $\theta_R$
which represent the mixing angles of the $b_L$ and $b_R$ with $D_L$
and $D_R$ respectively. Expressing the gauge eigenstates for the
mixing quarks as $b^0$ and $D^0$, the mass matrix in the ($b^0,D^0$)
basis is given by
\begin{align} \label{bDmass}
\mathcal{M} &= \left(\begin{matrix}
 y_b \, v_H/\sqrt{2} &  0 \\
 Y_{Db} \, v_S/\sqrt{2} &  M_D
\end{matrix}
\right).
\end{align}
This matrix can be diagonalized with a bi-unitary transformation
$\mathcal{M}_{\text{diag}}=\mathcal{R}_L \mathcal{M} \mathcal{R}_R^\dag$, where $\mathcal{R}_L$ and $\mathcal{R}_R$
are unitary matrices which rotate the left-chiral and right-chiral gauge eigenstates
to the mass eigenstates respectively. The interaction of the physical mass eigenstates
($b,D$) can then be obtained by writing the gauge basis states as
\begin{align} \label{Dstates}
b_i^0 &=  b_i \cos\theta_i + D_i \sin\theta_i
, &
D_i^0 &= - b_i \sin\theta_i + D_i \cos\theta_i.
\end{align}
The rotation matrices $\mathcal{R}_i$ are given by
\begin{align} \label{Drot}
\begin{aligned}\mathcal{R}_i &=
\begin{pmatrix}
 \cos\theta_i &  \sin\theta_i \\
 -\sin\theta_i & \cos\theta_i
\end{pmatrix},&
&\textrm{where } i=L,R.
\end{aligned}
\end{align}
 The corresponding mixing angles for the left- and
right-handed fields follow from diagonalizing the matrices
$\mathcal{M}\mathcal{M}^{\dag}$ and $\mathcal{M}^{\dag}\mathcal{M}$
respectively and are given by
\begin{align} \label{Dmix}
\tan 2\theta_L &= \frac{-2 \, Y_{Db} \, y_b \, v_S \, v_H}{2M_D^2 + Y_{Db}^2 v_S^2 - y_b^2 v_H^2}  , &
\tan 2\theta_R &= \frac{-2\sqrt{2} \, Y_{Db} \,v_S \, M_D}{2M_D^2 - Y_{Db}^2 v_S^2 - y_b^2 v_H^2}.
\end{align}
\section{Phenomenological Implications} \label{sec:pheno}

In hadronic
colliders such as the LHC and Tevatron, the dominant signals arise from the pair productions of
the exotic colored quarks, $D$ and $\overline{D}$, and their
subsequent decays (because $D$ has hypercharge, the LEP2 bound of $\sim$
100 GeV on its mass applies \cite{lep2}). The other important production process
is the pair productions of the exotic quark in association with the new $U(1)^\prime$
gauge boson,  $D\overline{D}Z^\prime$. It turns out that this is the only
way the new gauge boson $\Zp$ can be produced on-shell at LHC because of its
very suppressed or vanishing couplings to the SM particles in this model.
In the following subsections we discuss the signals from the $D\overline{D}$
production. We also discuss the couplings
of the extra gauge boson $Z^\prime$ with the SM particles.

\subsection{Signals from $D\overline{D}$ Productions}

\begin{figure}[tb]
\begin{center}
\includegraphics[width=4in]{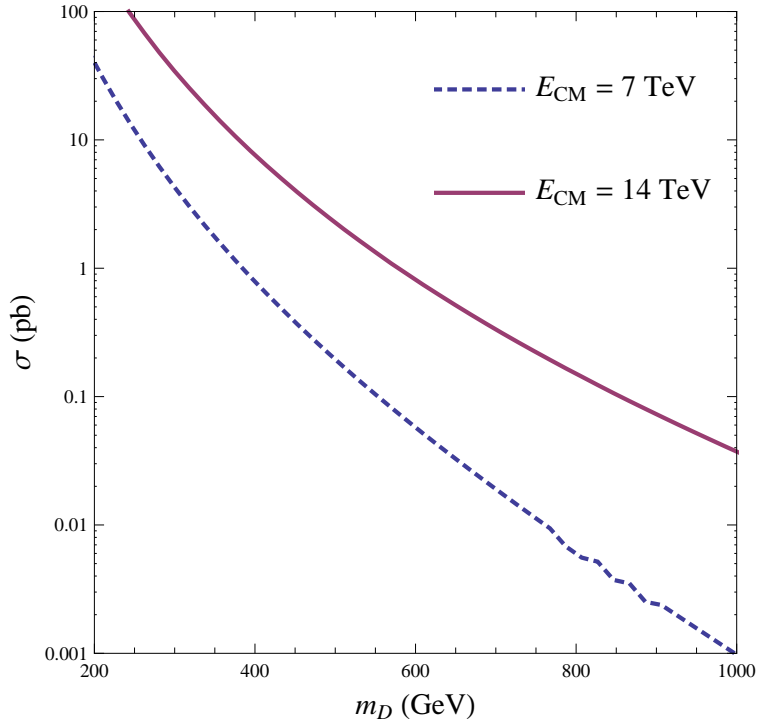}
\caption{\textsl{Pair production cross section for the exotic quarks at LHC as a function of
its mass ($m_D$). We use the CTEQ6L1 parton distribution functions (PDF) \cite{cteq6l} for the protons.
We have set the scale $Q^2=m_D^2$.}}
\label{fig:csec}
\end{center}
\end{figure}

The heavy exotic quarks being colored particles will be produced
copiously at the LHC through strong interactions. The major
contribution, as in the case of top quarks, would come from the
gluon induced subprocess ($\sim$ 80\%). In Fig.~\ref{fig:csec} we
plot the pair production cross section for the process
\begin{align}
p p &\longrightarrow D \overline{D}
\end{align}
at LHC for two different center-of-mass energies (7 TeV and 14 TeV).
%%%%%%%%%%%%
%%%%%%%%%%%%
The figure clearly shows that one can have quite large production
cross section for such an exotic quark at the LHC and its signals
should be observable through its decay products. We have implemented
the model into CalcHEP \cite{calchep} to calculate the production
cross sections as well as the two-body decays of the new particles in the model.

The heavy quark in the gauge eigenbasis couples directly to the
$Z^\prime$ gauge boson through the $U(1)^\prime$ charge, with the
gauge coupling strength $g_{\zp}$. However, its decay is more
dependent on the mixing parameters resulting from its mixing with
the $b$ quark, leading to a much richer phenomenology. The heavy
exotic itself is a mixed mass eigenstate and we list its couplings
to the other particles of the model in Table \ref{feynrules}.
\renewcommand{\arraystretch}{1.2}
\begin{table}[!p]
\begin{center}
\begin{tabular}{|l|c|c|c|}\hline
                            &$K$                                        &$c_V$                                  &$c_A$\\\hline\hline
$\overline{b}b Z_\mu$        &$\dfrac{e}{12\sin2\theta_W}$ &$4\cos2\theta_W+3\cos2\theta_L-1$  &$3(1+\cos2\theta_L)$\\\hline
$\overline{D}D Z_\mu$        &$\dfrac{e}{12\sin2\theta_W}$ &$4\cos2\theta_W-3\cos2\theta_L-1$  &$3(1-\cos2\theta_L)$\\\hline
$\overline{b}D Z_\mu$        &$\dfrac{-e}{4}\dfrac{\sin2\theta_L}{\sin2\theta_W}$    &1&1\\\hline
$\overline{b}b Z^\prime_\mu$ &$\dfrac{g_{\zp}Y_{\zp}}{4}$  &$\cos2\theta_L+\cos2\theta_R-2$    &$\cos2\theta_L-\cos2\theta_R$\\\hline
$\overline{D}D Z^\prime_\mu$ &$\dfrac{-g_{\zp}Y_{\zp}}{4}$ &$\cos2\theta_L+\cos2\theta_R+2$    &$\cos2\theta_L-\cos2\theta_R$\\\hline
$\overline{b}D Z^\prime_\mu$ &$\dfrac{-g_{\zp}Y_{\zp}}{4}$ &$\sin2\theta_L+\sin2\theta_R$&$\sin2\theta_L-\sin2\theta_R$\\\hline
$\overline{t}b~W^+_\mu$      &$\dfrac{-e\cos\theta_L}{2\sqrt{2}\sin\theta_W}$   &1&1\\\hline
$\overline{t}D~W^+_\mu$      &$\dfrac{e\sin\theta_L}{2\sqrt{2}\sin\theta_W}$   &1&1\\\hline\hline
                             &$K$                                        &$c_S$                                  &$c_P$\\\hline\hline
$\overline{b}b \phi_H $        &   $\dfrac{\cos\theta_R}{\sqrt{2}}$  & $y_{b}\cos\beta\cos\theta_L+y_{Db}\sin\beta\sin\theta_L$ &0\\\hline
$\overline{D}D \phi_H $        &   $\dfrac{\sin\theta_R}{\sqrt{2}}$  & $y_{b}\cos\beta\sin\theta_L-y_{Db}\sin\beta\cos\theta_L$ &0\\\hline
$\overline{b}b \phi_S $        &   $\dfrac{-\cos\theta_R}{\sqrt{2}}$ & $y_{b}\sin\beta\cos\theta_L-y_{Db}\cos\beta\sin\theta_L$ &0\\\hline
$\overline{D}D \phi_S $        &   $\dfrac{-\sin\theta_R}{\sqrt{2}}$ & $y_{b}\sin\beta\sin\theta_L+y_{Db}\cos\beta\cos\theta_L$ &0\\\hline
$\overline{b}D \phi_H$  &$\dfrac{1}{2\sqrt{2}}$&$\begin{matrix}y_{Db}\sin\beta\cos(\theta_L+\theta_R)~~\\~~-y_{b}\cos\beta\sin(\theta_L+\theta_R)\end{matrix}$&$\begin{matrix}y_{Db}\sin\beta\cos(\theta_L-\theta_R)~~\\~~-y_{b}\cos\beta\sin(\theta_L-\theta_R)\end{matrix}$\\\hline
$\overline{b}D \phi_S$  &$\dfrac{1}{2\sqrt{2}}$&$\begin{matrix}y_{Db}\cos\beta\cos(\theta_L+\theta_R)~~\\~~+y_{b}\sin\beta\sin(\theta_L+\theta_R)\end{matrix}$&$\begin{matrix}y_{Db}\cos\beta\cos(\theta_L-\theta_R)~~\\~~+y_{b}\sin\beta\sin(\theta_L-\theta_R)\end{matrix}$\\\hline
\end{tabular}
\caption{\small\textsl{The effective coupling of the exotic $D$ quark with the other particles in the model. The electromagnetic coupling with photon and the strong coupling
with gluon is the same as any down-type quarks in the SM.  Couplings are of the form $K\gamma^\mu(c_V-c_A\gamma^5)$ and $K(c_S-c_P\gamma^5)$. Note that we have put
$V_{tb}=1$.}} \label{feynrules}
\end{center}
\end{table}

\begin{figure}[!t]
\begin{center}
\begin{tabular}{cc}
\includegraphics[width=2.9in]{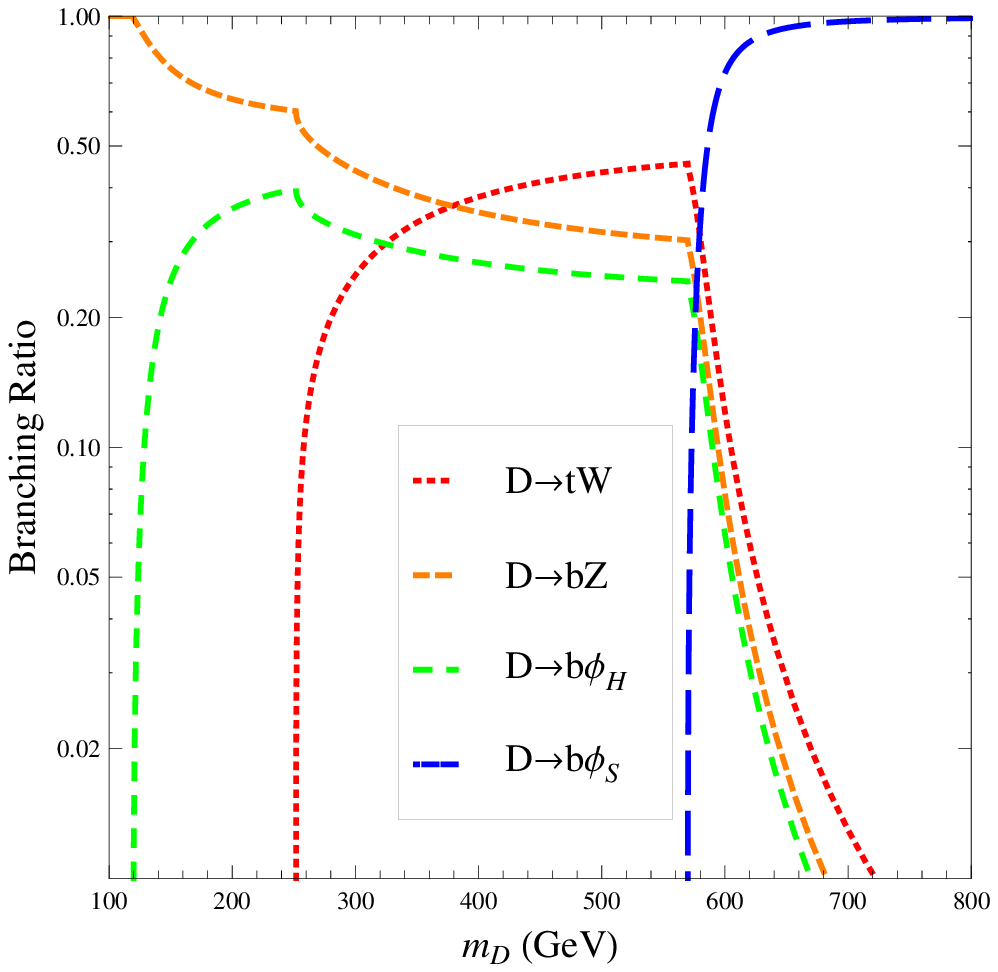}&
\includegraphics[width=2.9in]{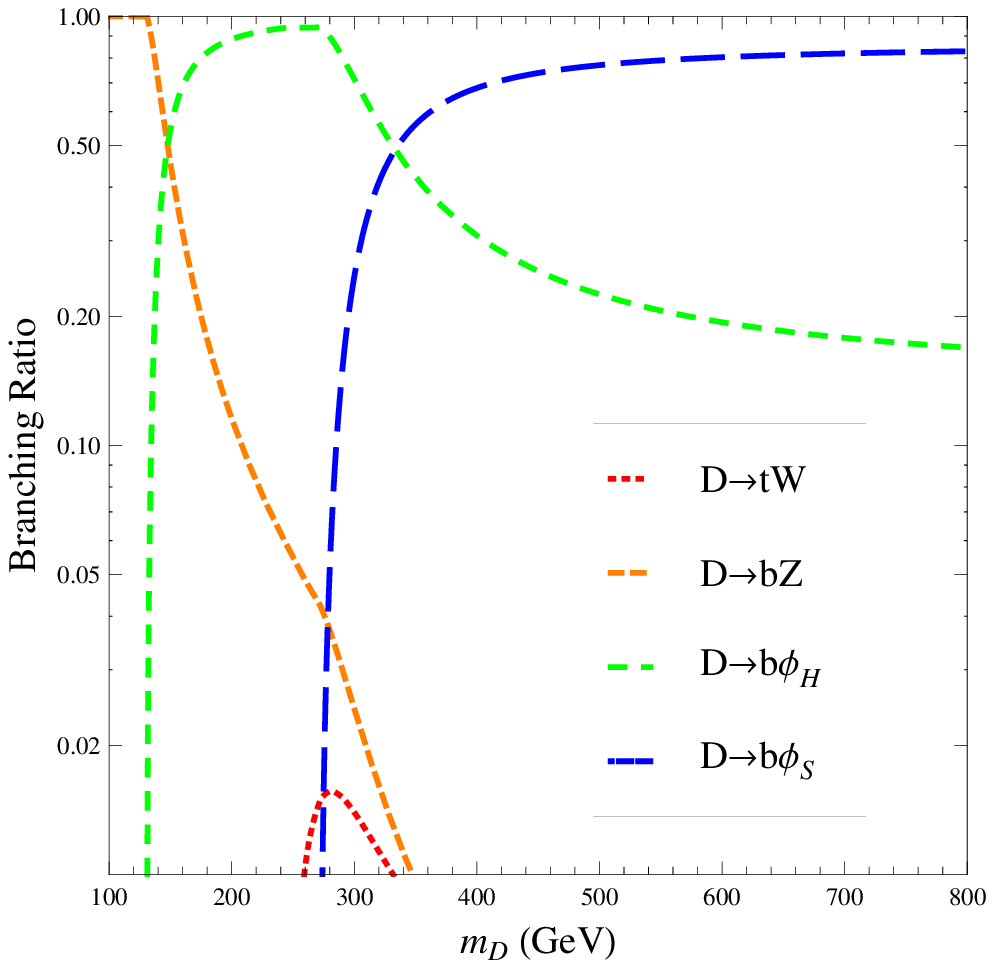}\\
\textbf{(a) Parameter Set I}&
\textbf{(b) Parameter Set II}
\end{tabular}
\caption{\textsl{Illustrating the decay probabilities of the $D$ quark as
a function of its mass $m_D$.}}\label{fig:bratio}
\end{center}
\end{figure}

The two body decay width for $D$ of mass $m_D$, in its rest frame can be written down as
\begin{align}
\Gamma(D\rightarrow X_2 X_3)&=\dfrac{1}{16\pi m_D} \lambda^{1/2} \left(1,\frac{m_{X_2}^2}{m_D^2},\frac{m_{X_3}^2}{m_D^2}\right)~~\overline{|\mathcal{M}|^2}
\end{align}
where the function $\lambda(x,y,z)=x^2+y^2+z^2-2(xy+yz+zx)$.
Using the effective couplings given in Table~\ref{feynrules}, one can write down the explicit decay amplitudes for the exotic quarks decaying into
vector ($V$) and scalar ($\Phi$) modes.
\begin{align}
\begin{split}
\overline{|\mathcal{M}|^2}(D\rightarrow fV) &= K^2\left[ 3\left(m_D^2+m_f^2-2m_V^2+\dfrac{(m_D^2-m_f^2)^2}{m_V^2}\right)(c_V^2+c_A^2) \right. \\
&\qquad \left. -18m_D m_f(c_V^2-c_A^2) \right]  \\
\overline{|\mathcal{M}|^2}(D\rightarrow f\Phi) &= K^2\left[ 3\left(m_D^2+m_f^2-m_\Phi^2 \right)(c_S^2+c_P^2) + 6 m_D m_f(c_S^2-c_P^2)\right]
\end{split}
\end{align}
We can now estimate the decay probabilities of the heavy exotic $D$
quark. To highlight distinct scenarios, we choose two different sets
of input values for the free parameters as representative points in
the model listed in Table~\ref{parameters}. Note that the input
parameters for the model also affect some EW observables, e.g.~the
$Z$ boson decay width or the mass limits for Higgs boson and other
heavy exotics that appear in our model.
%%%
%\renewcommand{\arraystretch}{1.0}
\begin{table}[!t]
\begin{center}
\begin{tabular}{|c||c|c|}
\hline {\bf Parameters} & {\bf I} & {\bf II} \\ \hline
\hline $(\lambda_H,\lambda_S,\lambda_{HS})$ & $(0.11,0.16,0.005)$ & $(0.2,0.05,0.1)$ \\
\hline $v_S$  & $1000$ GeV & $800$ GeV \\
\hline $Y_{Db}$ & $0.15$ & $0.05$ \\ \hline
\hline $m_{\phi_H}$  & $115$ GeV & $127$ GeV \\
\hline $m_{\phi_S}$  & $566$ GeV & $268$ GeV \\
\hline $m_{\Zp}$  & $1000$ GeV & $800$ GeV \\ \hline
\end{tabular}
\caption{\textsl{Representative points in the model parameter space and
the relevant mass spectrum used in the analysis.}}\label{parameters}
\end{center}
\end{table}
%\noindent
We have checked that the input parameters given in Table~\ref{parameters} are allowed and do not
contradict any existing experimental bounds \cite{lep2}. In Fig.~\ref{fig:bratio} we present the decay branching ratios (BR) of the heavy quark $D$ as a function of its mass ($m_D$) for the
representative points I \& II given in Table~\ref{parameters}.
%%%%%%%%%%%%
%%%%%%%%%%%%
%\FIXME{Can we make a branching ratio graph for $Z^\prime$ and $\phi_S$?}
The curves in Fig.~\ref{fig:bratio}(a) represent Point-I from
Table~\ref{parameters}. When $D$ is lighter than $m_t+M_W$ then it
always decays to $Z \, b$ through mixing if its coupling to the lighter
Higgs boson is very suppressed. This would happen when the lighter
scalar state is dominantly an $SU(2)$ doublet. The $t \, W^-$ mode
starts picking up and becomes comparable to the $Z \, b$ mode for
heavier $D$.  $t \,W^-$ is a common decay mode in 4th-generation models
and theories with top or bottom partners as studied in
Ref.\cite{Contino:2008hi} and results in multi-lepton signals.
% Thus the
%most interesting decay modes for us are $Z \, b$ and $t \, W^-$.

The curves in Fig.~\ref{fig:bratio}(b) represent Point-II, where the choice of parameters
give a very suppressed mixing angle $\theta_L$. The couplings of $Z \, b$ and
$t \, W$ to the exotic quark are proportional to $\sin\theta_L$ and hence
also get suppressed. As soon as the scalar modes become kinematically accessible,
they completely dominate the decay properties of the exotic quark.

\begin{figure}[!ht]
\begin{center}
\begin{tabular}{cc}
\includegraphics[width=2.9in]{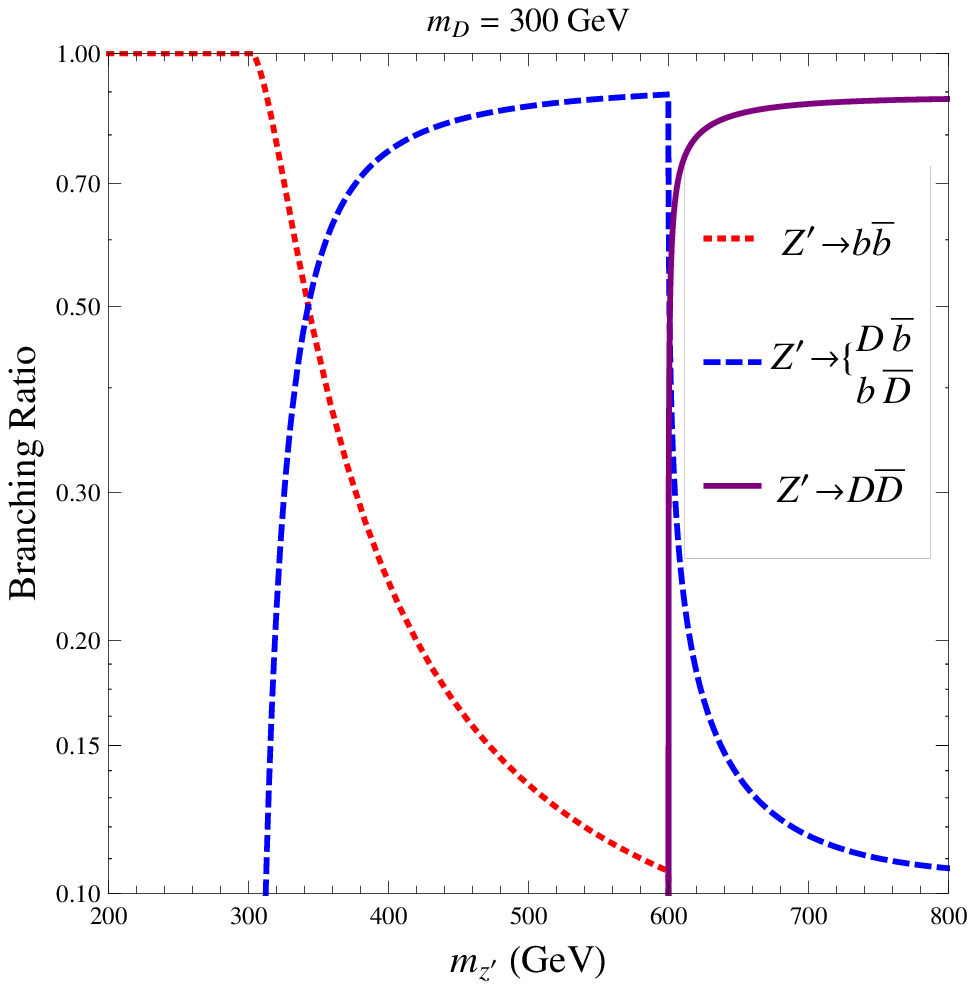}&
\includegraphics[width=2.9in]{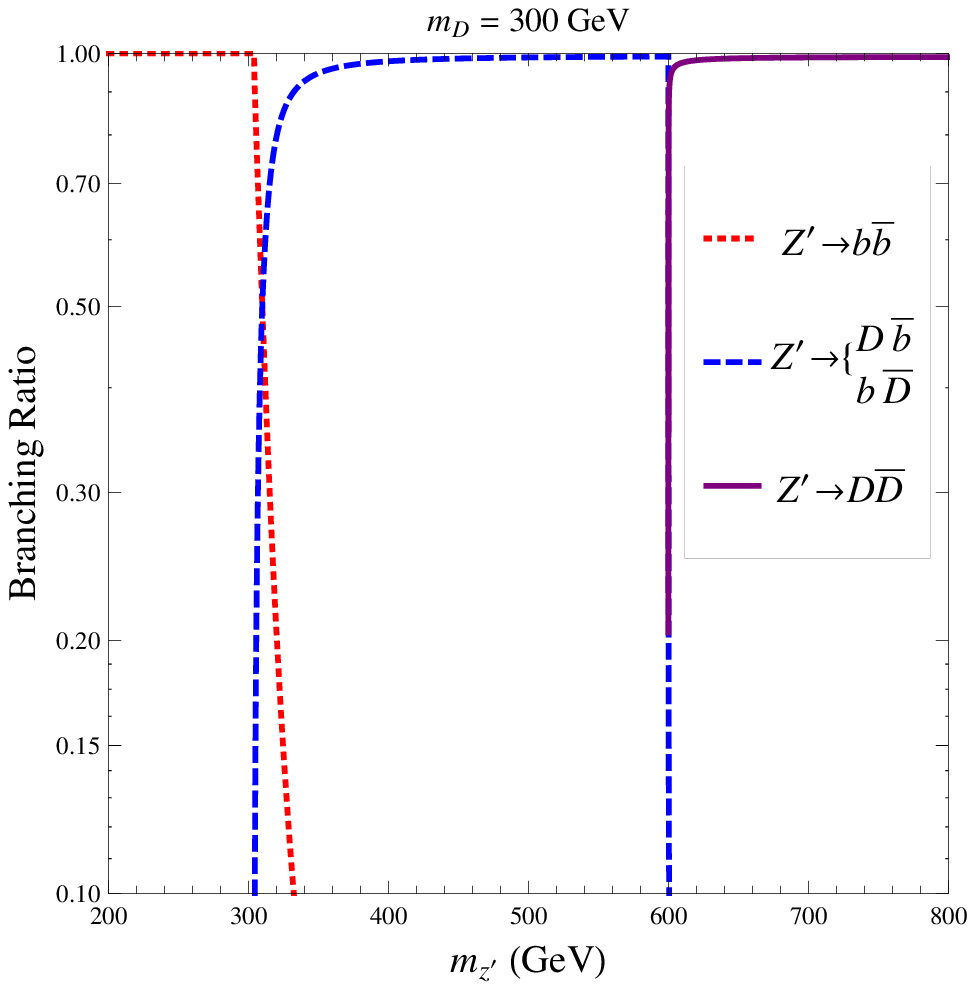}\\
\textbf{(a) Parameter Set I}&
\textbf{(b) Parameter Set II}
\end{tabular}
\caption{\textsl{Illustrating the decay probabilities of the $Z^\prime$ as
a function of its mass $m_{Z^\prime}$.}}
\label{Zprimedk}
\end{center}
\end{figure}

%%%%%
\begin{table}[!b]
\begin{center}
\begin{tabular}{|l|cc|l|cc|} \hline
 &\multicolumn{2}{|c|}{Branching Ratios}& &\multicolumn{2}{|c|}{Branching Ratios}\\ \cline{2-3}\cline{5-6}
\multicolumn{1}{|c|}{Decays}    &  I    &  II
&\multicolumn{1}{|c|}{Decays}    &   I   &  II  \\ \hline $\phi_H
\to b\bar{b}$   & 0.672 &  0.510 &  $\phi_S \to \phi_H \phi_H$  &
0.25 & 0.27 \\ \hline $\phi_H \to c\bar{c}$   & 0.031 &  0.024 &
$\phi_S \to W^+W^-$         &  0.42 & 0.51 \\ \hline
$\phi_H\to\tau^+\tau^-$ & 0.093 &  0.072 &  $\phi_S \to Z Z$ &  0.20
& 0.22 \\ \hline $\phi_H \to gg      $   & 0.104 &  0.096 & $\phi_S
\to t \bar{t}$      &  0.13 & -- \\ \hline $\phi_H \to W W^*$      &
0.088 &  0.266 &   \multicolumn{3}{c}{}    \\\cline{1-3}
\end{tabular}
\caption{\textsl{Branching Ratios for various Higgs decay modes for
parameter sets I and II.}} \label{higgsdk}
%\vspace*{0.1in}
\end{center}
\end{table}

We find if the $Z^\prime$ boson is light, then as soon as the
$D\to\Zp b$ mode opens up, the remaining modes drop out very quickly
for Point-I while for Point-II it becomes comparable to the scalar
mode for very large mass $m_D$. It is worth pointing out here that
the dominant decay mode for the $Z^\prime$ when $m_{Z^\prime} < m_D$
is to $b\bar{b}$ with 100\% branching probability. However for $\Zp$
heavier than $D$ the dominant decay of $\Zp$ is to $D\bar{b}$ and
$\bar{D}b$ as shown in Fig.~\ref{Zprimedk}; and as soon as
$m_{Z^\prime} > 2\ m_D$ it decays dominantly to $\bar{D}D$ with
maximum probability. We should also point out that the $\Zp$
phenomenology in our model is quite different from other models with
$U(1)$ extension of the SM. As there exists no coupling between any
SM fermion pair (other than $b$ quark) or with the EW gauge bosons
(no kinetic mixing), it is not possible to produce this particle
directly through exchange of SM particles at LEP or Tevatron and so
the strong constraints that exist on the mass of similar $\Zp$
exotics through the effective four-fermion operators
\cite{langacker} do not apply in our case and neither do the search
limits from the Tevatron experiments \cite{tevatron}. Thus the $\Zp$
in our model can be light but remain invisible in the existing
experimental data. We will however not discuss the $\Zp$ signals any
further and only focus on the signals arising from the production of
the exotic $D$ quarks in the model. To understand the full decay
chain of the $D$ quark to final state particles we also list the
decay probabilities of the scalars $\phi_H$ and $\phi_S$ in
Table~\ref{higgsdk} for the two representative points I \& II.

Thus the above decay patterns suggest that one can have the
following interesting final states from the decay of the exotic
quarks
\begin{align}
p p \rightarrow D\overline{D} &\rightarrow
\left\{\begin{aligned}
&(Z b)(Z \bar{b})           &&\Longrightarrow b\bar{b}+2 Z      \\
&(t W^-)(\bar{t} W^+)       &&\Longrightarrow t\bar{t}+W^+W^-   \\
&(t W^-)(Z \bar{b})         &&\Longrightarrow t\bar{b}+ Z W^-   \\
&(\phi_H b)(\phi_H \bar{b}) &&\Longrightarrow b\bar{b}+2 \phi_H         &&\to b\bar{b} + 2(W^+ W^-) \\
&(\phi_H b)(\phi_H \bar{b}) &&\Longrightarrow b\bar{b}+2 \phi_H         &&\to 3 (b\bar{b})\\
&(\phi_S b)(\phi_S \bar{b}) &&\Longrightarrow b\bar{b}+2 \phi_S         &&\to 3 (b\bar{b}) \\
&(\phi_H b)(\phi_S \bar{b}) &&\Longrightarrow b\bar{b}+\phi_H \phi_S    &&\to b\bar{b} + 3\phi_H    &&\to 4 (b\bar{b}) \\
&(\phi_S b)(\phi_S \bar{b}) &&\Longrightarrow b\bar{b}+2 \phi_S         &&\to b\bar{b} + 4 \phi_H   &&\to 5 (b\bar{b})
\end{aligned}\right.
\end{align}
where the first four suggest multi-lepton and multi-jet final
states with two or more $b$-jets, while the remaining give more exotic
signatures like  $N$ $b$-jet final states where $N$ can be as large as
10. Note that the above decays only illustrate some of the possible
decay chains and we have not listed other possible combinations of
the $D$ decays which can also lead to similar final states.

In Table~\ref{branchings} we list the probabilities for the decay
modes for a few specific values of the $D$ quark mass. We also show
the corresponding cross sections for the pair production of these
exotics at LHC for the center of mass energies of $\sqrt{s}=7$ TeV
and $\sqrt{s}=14$ TeV. The decays suggest a large multiplicity of
$b$ quarks in the final state. It turns out that six-$b$ final
states for the signal is very promising. However, there exists no
estimate for this  final state in the literature, arising from the
SM. We present below a leading-order (LO) estimate of the cross
section for the six-$b$ SM background from QCD for LHC energies.
%%%%%%%
\begin{table}[!t]
\begin{center}
\begin{tabular}{|c|c|c|cc|cc|cc|cc|}  \cline{4-11}
\multicolumn{3}{c}{}&\multicolumn{8}{|c|}{Branching Ratios}\\\hline
\multicolumn{1}{|c|}{\multirow{2}{*}{$m_D$}} &\multicolumn{2}{|c|}{$\sigma(D\bar{D})$(pb)} & \multicolumn{2}{|c|}{$D\to tW$} &  \multicolumn{2}{|c|}{$D\to bZ$} &  \multicolumn{2}{|c|}{$D\to b\phi_H$} &  \multicolumn{2}{|c|}{$D\to b\phi_S$} \\\cline{2-11}
    & 7 TeV & 14 TeV & I & II & I & II & I & II & I & II \\ \hline
250 & 12.15 & 87.760 & 0     & 0     & 0.603 & 0.055 & 0.397 & 0.945 & 0     & 0    \\
300 & 4.265 & 34.368 & 0.251 & 0.014 & 0.438 & 0.024 & 0.311 & 0.715 & 0     & 0.247\\
400 & 0.791 &  7.692 & 0.381 & 0.005 & 0.351 & 0.005 & 0.268 & 0.308 & 0     & 0.681\\
500 & 0.194 &  2.270 & 0.434 & 0.003 & 0.316 & 0.002 & 0.250 & 0.225 & 0     & 0.770\\
600 & 0.059 &  0.820 & 0.121 & 0.002 & 0.078 & 0.001 & 0.063 & 0.194 & 0.738 & 0.803\\\hline
\end{tabular}
\caption{\textsl{Cross sections and branching probabilities for specific mass values of $D$ quark
for the representative points I and II.}}
\label{branchings}
\end{center}
\end{table}
%%%%%
\subsection{Calculation of Six-$b$ Final States from QCD}
%%%
The six-$b$ final state is interesting, independent of our
particular model.  The presence of six $b$ jets allow the jets to be
tagged.  All other 6-jet final states involve mixtures of light
quarks and gluons, and one cannot separate light jets from gluon
jets.  Therefore the six-$b$ final state itself presents an
interesting test of QCD.  Furthermore by computing the full Matrix
Element, we can test the validity of the differential cross section
by looking at differential observables. While we only use the
six-$b$ cross section as a background to our signal, this is the
first time such a six-$b$ cross section in QCD has been estimated,
and this also is an important result of this paper.

% our
%contributions to MadGraph enabling the computation of higher
%multiplicity of QCD final states is also one of our major results of
%this paper. \CRed{(Are we going to contribute back to Madgraph at
%this stage? If not then the last sentence needs to change.)}

Any six-final state process is a challenge to compute. The phase space
is 20 dimensional, there are thousands of diagrams, and thousands of
distinct color configurations.  While six-jet final states are produced
every day by Monte Carlo generators such as PYTHIA \cite{pythia}
and HERWIG \cite{herwig}, the mechanism they use creates additional jets from
an initial $2 \to 2$ or $2 \to 3$ process via a showering procedure that
resums leading logarithms, splitting an extra gluon from the hard final
state partons.

As is well known, the showering procedure cannot produce the correct
correlations among 3 or more hard jets, nor can it compute the total
cross section for 3 or more hard jets.  It assumes that each parton is
independent of all the others.  For a single radiation it is strictly
correct in the limit that the extra radiation is soft and/or collinear
with the inital parton.  However for multiple radiations it ignores the
QCD connection among the radiations, assuming that each radiation
factorizes from the others.  There is also quantum mechanical
interference in different radiations which result in the same final
state that is ignored.

This means that one should not examine in detail observables such as the
angles between jets, invariant masses of jet pairs, or the thrust, when
one hard jet came from the showering procedure.  This showering
technique is however extremely useful as long as one is not sensitive to
the details of correlations in the differential cross section, as this
method is computationally simpler than a full Matrix Element
calculation.

Therefore to have an accurate Monte Carlo with six jets in the final
state, one must compute the full Matrix Element, which automatically
includes all color flows and interference.  This is accurate to
approximately the 10\% level, at which point NLO loop corrections become
important.  Note that due to the $b$ quark mass, there is no soft
radiation which benefits significantly from the usual Sudakov logarithm
resummation.  The $b$ mass acts as a regulator, relegating this cross
section strictly into the ``hard'' regime, in which the Matrix Element
is valid.  Even if all six $b$'s are at rest, the energy in the final
state is 30 GeV, and any virtual gluon must have a virtuality $q^2 \sim
10$ GeV.

For this calculation we have chosen the tool MadEvent, which is an
event generator built upon the Matrix Element generator MadGraph
\cite{madgraph}. We have modified these tools to be able to cope
with thousands of diagrams and thousands of color configurations.
Computing 5 and 6 final state QCD processes has a number of
challenges, all of which are technical rather than physics-based.
MadEvent is in principle capable of computing any process with any
number of final state particles, however several internal
restrictions caused previous versions to fail on processes such as 6
$b$'s. Although the program is equipped to handle the large color
configurations, some input/output statements restrict this
computation to a smaller number of color flow configurations which
makes it incapable of calculating the six $b$ final states at a
hadron machine like the LHC. These have been repaired to calculate
the six $b$ cross section in the SM from QCD at the LHC.

\subsection{Signal and Background Analysis}
%\FIXME{We need some histograms in this section}
A simple minded estimate of the cross section using $\sigma\times BR$ shows that the
final states which would be of interest at the LHC would involve at least 2 $b$-jets
in the final state. Besides the two hard $b$-jets, one expects charged leptons in the final
state coming from the decays of the weak gauge bosons. It is also worth noting that when
the $D$ quark decays to the Higgs bosons, one would get a large multiplicity of $b$-jets in
the final states as the Higgs with $M_{\phi_H} < 2 M_W$ dominantly decays to $b$-jets.
To select our events for the final states given in Table~\ref{cross-sections}, we
have imposed the following kinematic cuts:
\begin{itemize}
\item All the $b$-jets must have a $p_T^{b} > 20$ GeV and lie within the rapidity gap
of $|\eta^{b}|<3.0$.
\item All charged leptons ($\ell=e,\mu$) must have a $p_T^{\ell} > 20$ GeV and lie
within the rapidity gap of $|\eta^{\ell}|<2.5$.
\item The final states also must satisfy $\Delta R_{bb}>0.7$, $\Delta
R_{b\ell}>0.4$, and $\Delta R_{\ell\ell}>0.2$ where $\Delta R_{ij} =
\sqrt{(\Delta\eta_{ij})^2 + (\Delta \phi_{ij})^2}$.
\item All $b$-jet pairs must have a minimum invariant mass $M_{bb}> 10$ GeV.
\end{itemize}
In Table~\ref{cross-sections} we present the cross-sections for the
signal for two different mass values of the exotic $D$ quark after
passing through the above mentioned kinematic cuts. As expected, the
favored final states are dependent on the high $b$-jet multiplicity.
At the hadron collider such as LHC, one favors final states with
leptons. However $b$-jets can also be triggered upon and identified
and thus can prove to be useful in isolating new physics signals
such as ours which involve at least two or more $b$-jets. Looking at
Table~\ref{cross-sections} we find that we get a good signal rate
for the inclusive $6b+X$ final state. The SM background for
multi-$b$ final state is quite large \cite{baer}. However no
estimate of a $6b$ final state exists in the literature, which we
find relevant for our signal. We have used the {\it
Madgraph+MadEvent} package to estimate the leading
order partonic cross section for the $6b$ final state at LHC. With the above
mentioned kinematic cuts, we find that for LHC energy of
$\sqrt{s}=14$ TeV, the SM background for $6b$ final state is $\sim
70$ fb and falls to less than $10$ fb for the $\sqrt{s}=7$ TeV
option. This implies that the signal in our model is much larger
than the SM background even for larger mass values of the exotic $D$
quark.  The other signals which are worth looking for in this model
is one or two charged leptons with varying $b$-jet multiplicities.
We have listed the interesting ones in Table~\ref{cross-sections}.
The
%%%%%%%%%%%%%%%%%%%%%%
\begin{table}[!t]
\begin{center}
\begin{tabular}{c|c|c|c|c|c|c|c|c|} \cline{2-9}
    &\multicolumn{4}{|c|}{$m_D=300$ GeV} & \multicolumn{4}{|c|}{$m_D=500$ GeV}\\\hline
\multicolumn{1}{|c|}{\bf{Final}}&\multicolumn{2}{|c|}{$\sqrt{s} = 7$ TeV}&\multicolumn{2}{|c|}{$\sqrt{s} = 14$ TeV}&\multicolumn{2}{|c|}{$\sqrt{s} = 7$ TeV} & \multicolumn{2}{|c|}{$\sqrt{s} = 14$ TeV}\\ \cline{2-9}
\multicolumn{1}{|c|}{\bf{States}}&  I    &   II  &  I     &     II&  I    &   II  &  I     &     II\\ \hline
\multicolumn{1}{|c|}{$6b+X$}    &181.92&718.79&1394.32&5521.23&4.94&10.79&531.10&115.02\\\hline
\multicolumn{1}{|c|}{$nb+\ell+X$ ($n\ge 2$)} &452.50&188.22&3559.94&1465.42&32.17&27.88&43.37&313.96\\\hline
\multicolumn{1}{|c|}{$2b+2\ell+X$} &146.53&14.95&1127.15&117.08&8.71&6.61&11.77&75.56\\\hline
\multicolumn{1}{|c|}{$4b+2\ell+X$} &51.07&24.36&384.24&183.73&1.85&2.06&20.33&22.20\\\hline
\end{tabular}
\end{center}
\caption{\textsl{Illustrating the final state cross sections after the decay of
$D$ quarks. All cross sections are in units of femtobarn (fb).}}
\label{cross-sections}
\end{table}
%%%%%%%%%%%%%%%%%%%%%%%
final state with $2\ell+4b+X$ also stands out against the SM background, where one
gets the SM cross sections to be quite small as it is already $\alpha_{EW}/\alpha_s$
suppressed compared to the $6b$ cross section. The SM background is much larger for the
final states $\ell+nb+X$ where $n=2$ and $2\ell+2b+X$, where the significant
SM background results from the $t\bar{t}$ production. For the
final states $\ell+nb+X$ one can get rid of the huge $t\bar{t}$ background by
demanding $n \ge 3$. This helps in improving the significance of the signal, even though
we also lose a large fraction of the signal events in the process. For the other final
state, we find that at leading order, at LHC with $pp$ collision energy of
$\sqrt{s}=7$ TeV, the  $2\ell+2b+X$ SM background is
$\sim 3.3$ pb. As the leading two $b$-jets in our signal come from the decays
of the heavy exotic $D$ quark, we put a stronger $p_T$ cut of $100$ GeV. This
reduces the signal by two-thirds. However the SM background is
reduced by more than an order of magnitude, and becomes $\sim 232$ fb for
$\sqrt{s}=7$ TeV collisions while it is $\sim 1.63$ pb for $\sqrt{s}=14$ TeV which
does look promising for the signal with large enough luminosities at the LHC. We must
point out that we have not incorporated any efficiency
factors for our final state particles. Most notably, all numerical estimates involving
$b$-jets for signal as well as the SM background will have to be scaled with the
$b$-tagging efficiency of around 50-60\% expected at the LHC \cite{Tomalin:2007zz}.
%\FIXME{Include efficiency factors\dots}

%\begin{table}[!t]
%\begin{center}
%\begin{tabular}{c|c|c|c|c|}\cline{1-5}
%\multicolumn{1}{|c|}{\bf{Final}}&\multicolumn{2}{|c|}{without cuts}&\multicolumn{2}{|c|}{with cuts}\\\cline{2-5}
%\multicolumn{1}{|c|}{\bf{States}}&$\sqrt{s}=7$ TeV&$\sqrt{s}=14$ TeV&$\sqrt{s}=7$ TeV&$\sqrt{s}=14$ TeV\\\hline
%\multicolumn{1}{|c|}{$6b+X$}    &   &82.87 pb &   & 70.33 fb   \\\hline
%\multicolumn{1}{|c|}{$Nb+\ell+X$ ($n\ge 2$)}  &   &       &   &   \\\hline
%\multicolumn{1}{|c|}{$2b+2\ell+X$} &   &       & 3.32 (0.242) pb  &   \\\hline
%\multicolumn{1}{|c|}{$4b+2\ell+X$} &   &       &   &   \\\hline
%\end{tabular}
%\end{center}
%\caption{\sl Approximate values for the SM background for the different final states considered. We have
% used the package Madgraph+MadEvent \cite{mad} to estimate the background at parton level.}
%\end{table}

\section{Summary and Conclusions}

In this work, we have proposed a new extension of the SM, by
introducing a hidden $U(1)^{\prime}$ symmetry. The difference with
the previously studied $U(1)^{\prime}$'s is that all the SM
particles are singlets under our proposed new $U(1)^{\prime}$, and
hence hidden. Such a symmetry may be present at the TeV scale, and
may manifest at the LHC giving new signals observable at the LHC.
The model incorporates a new EW singlet Higgs, as well as new
vector-like charge $-1/3$ quarks. We have studied the pair
productions of these new quarks and their subsequent decays. The
dominant final states include  multiple $b$ jets with high $p_T$, or
$b$ jets plus charged leptons with high $p_T$ and missing energy,
and stands out beyond the SM background. The most distinctive final
state signal is the $6b$ quark with high $p_T$ and no missing
energy. A lot of effort have been put in both for the ATLAS and CMS
detectors to improve the $b$-tagging efficiency. So the calculation
for this $6b$ final state is also of great importance in the SM, and
has not yet been calculated. We have calculated this $6b$ signal in
our model, and have also estimated the SM expectation using MadGraph
and MadEvent. We found that the signal in our model stands well
above that expected from the SM.

\section*{Acknowledgements}

We thank T.~Stelzer and F.~Maltoni for several useful communications
regarding the MadGraph and MadEvent package. S.~Nandi thanks the
Fermilab Theory Group for their warm hospitality and support from
their Summer Visitor Program during the completion of this work. S.
Nandi also thanks the CERN Theory Group for warm hospitality and
support while part of this work was done. S.~K.~Rai would like to
acknowledge and thank the Helsinki Institute of Physics for their
support and hospitality. The work of BNG, SN and SKR is supported in
part by the United States Department of Energy, Grant Numbers
DE-FG02-04ER41306 and DE-FG02-04ER46140.

 \end{document}